%%%%%  Manuscript Version of Talk given 
%%%%%   at GRG16 in Durban, South Africa, July 16, 2001
%%%%%  this one in TeX, 
%%%   created mid August, 2001
\magnification=\magstep1
\baselineskip=20 pt plus 1pt minus 1pt
\vsize=8.9truein
\hsize=6.8truein
\hfuzz=1.5pt
\footline={\ifnum\pageno=1 \hfil\else\centerline{\folio}\fi}

\def\hi{\hangindent=15pt}
\def\vs{\vskip 6truept}

\def\veject{\vfill\eject}
\def\no{\noindent}
\def\h{\textstyle {1 \over 2}}

\def\ss#1#2{\vs\noindent{\bf \llap{#1.~~}#2\par}}

\def\ssb#1{\itemitem{#1)~}}

\def\today{\number\day\space\ifcase\month\or January\or February\or 
March\or April\or May\or June\or July
\or August\or September\or October\or November\or December\fi, \number\year} 

\font\smaller=cmr9
\font\smbold=cmbx9
\font\smchapt=cmbx12
\font\abstrct=cmr8
\font\reglar=cmr10

\def\p{\partial}
\def\cite#1{{$^{#1}$}}
\def\nought{{{}_{_{\!0}}}}
\def\tfrac#1#2{{\textstyle{#1\over#2}}}

%Section that Provides Notation for Differential Forms and Vectors
\def\dform#1{\smash{\raise.5pt\hbox{$\mathop{#1}\limits_{
    \raise3pt\hbox{\smash{\hbox{$\scriptstyle\sim$}}}}$}}}
\def\dom{{\dform{\omega}}}
%% This generates the nice "H" for H-spaces, and HH-spaces
\def\hsp{{\hscript h}}
\font\hscript=eufb10 scaled \magstep 1  %A Fraktur font

\smchapt 
\centerline{ Equations for Complex-Valued, Twisting, Type N, Vacuum Solutions,}
\vskip-2pt
\centerline{with one or two Killing/homothetic vectors}
\smallskip
\reglar
\centerline{{\bf Daniel Finley}$^\dagger$}
\abstrct
\vskip-3pt
\centerline{ Dept. of Physics and Astronomy, University of 
New Mexico}
\vskip-6pt
\no
Albuquerque, N.M. 87131 U.S.A.\hfill finley@tagore.phys.unm.edu
\vs
{\narrower{\narrower 
\baselineskip=12pt
{\smbold Abstract:~~}
\hsp\hsp-spaces, i.e., complex spacetimes, 
of Petrov type N$\times$N are determined by a trio of pde's for 
two functions, $\lambda$ and $a$, of three independent variables
(and also two gauge functions, chosen to be two of the independent 
variables if one prefers).  As in common integrable systems, these form a second order, linear system for $\lambda$; howver, here 
the integrability conditions, involving $a$, are more complicated 
than is common.  Therefore, with the hope of finding new solutions,
these equations are now constrained to also admit both one and 
two homothetic or Killing vectors.  The case with one Killing and
one homothetic vector reduces these equations to two ode's for two
unknown functions of the one remaining variable.
\par
In addition, we also describe in detail the explicit forms of the 
metric, tetrad, connections, and curvature for twisting 
\hsp\hsp-spaces of Petrov type N$\times$N, modulo the determining
equations.  This simplifies considerably the process of obtaining 
these details cleanly from earlier articles on the subject, 
thus simplifying access to the research area.
\smallskip}}

\vskip12pt
\reglar
\ss{1}{Simpler Description of the Local Metric}
\medskip
The goal of understanding general classes of solutions of Petrov Type N, with 
non-zero twist, is one that is still not realized.  The use of \hsp\hsp-spaces 
to forge a different path toward this goal was developed to 
a reasonable form in 1992.\cite{1}  This work pushes that path a step further, 
in a direction that has interested the more standard analysis for some time:  to
look for solutions that admit one Killing and also one homothetic vector, to
 simplify the task.\cite{2}
\vs
A general \hsp\hsp-space is a complex-valued solution of the Einstein vacuum
field equations that admits (at least) one congruence of null strings, i.e., 
a foliation by completely null, totally geodesic two-dimensional 
surfaces.\cite{3}  Those solutions with 
algebraically-degenerate, real Petrov type have 
two distinct such congruences.  We describe them with 
coordinates $\{p,v,y,u\}$, where 
$p$ is an affine, null coordinate along one null string,
$v$ specifies local wave surfaces,
and $y$ and $u$ are transverse coordinates,
in those surfaces.  The metric is determined by 
$x$ and $\lambda$, functions of $\{v,y,u\}$, 
 which must satisfy three 
quasilinear pde's.  (Alternatively, one may reverse the 
roles of $v$ and $x$, choosing $\{x,y,u\}$ as independent variables, 
and treating $v=v(x,y,u)$ and
$F \equiv v_x\,\lambda[v(x,y,u),y,u]$ as dependent variables.)
  These three equations are most easily presented using 
a mixed (non-holonomic) basis for the derivatives in 
these three variables:
$$
\p_1 \equiv \p_v\;,\quad \p_2\equiv \p_y\;,
\quad \p_3\equiv \p_u + a\p_v, \hbox{~with~}
a\equiv -x_u/x_v\;,\eqno(1)$$
where $\p_2$ is the derivative with respect to $y$ in the 
$\{v,y,u\}$ coordinate system while $\p_3$ is the derivative with respect to 
$u$ in the $\{x,y,u\}$ coordinate system, i.e., $v_2 = 0$
and $x_3 = 0$.  This then implies the commutators
$$ [\p_1,\p_2] = 0\;,\quad [\p_1,\p_3] = a_1\,\p_1\;,\quad
[\p_2,\p_3] = a_2\,\p_1\;,\eqno(2)$$
where the function $a \equiv v_3$ determines the twist of 
the metric, it being proportional to $a_2 = v_{32} \propto
x_{23}$, which we insist remain nonzero.
\vs
The constraining pde's then have the following form in terms of $\lambda$ and $a$:
$$\eqalign{
\lambda_{22} = \Delta\lambda\;,\qquad
\lambda_{33} +& 2a_1\lambda_3 + a_{31}\lambda = \gamma\lambda\;,\cr
a_2(\lambda_{23}+\lambda_{32}) + & a_{22}\lambda_3+a_{32}\lambda_2
+ \h a_{322}\lambda = 0\;,\cr}\eqno(3)$$
where 
two gauge functions, $\Delta$ and $\gamma$, have also been 
introduced that determine the left- and 
right-curvatures:
$$\eqalign{\Delta = \Delta(x,y) \hbox{~~so that~}
&\Delta_3 = 0\,,\hbox{~while~} C^{(1)} \propto \Delta_1 \ne 0 \;,\cr
\gamma= \gamma(v,u) \hbox{~~so that~}
&\gamma_2 = 0\,,\hbox{~while~} {\overline C}{}^{(1)}
\propto \gamma_1\ne 0\;.\cr}\eqno(4)$$
These functions do simply describe some gauge 
freedom in the defining equations, since they 
may be chosen arbitrarily, modulo the constraints 
above, as explained in more detail below.  
\vs
These constraints are sufficient to completely satisfy Einstein's equations.  
However, for them to be {\bf in involution}, 
we must also differentiate them, and solve, to 
determine either\cite{4} 
\ssb{a} $\lambda_{22}$, $\lambda_{33}$, $a_{322}$, and also $\p_3\lambda
_{22}$, $\p_2\lambda_{33}$, and $\p_1a_{322}$, {\bf ~or}
\ssb{b} all {\bf six} $\lambda_{ij}$, along with {\bf eight} equations 
between the three $\lambda_i$ which involve derivatives of 
$a$ up through the 6th-order. 
\vskip12pt
As (the only known) explicit 
example, Hauser's solution\cite5 is given by 
$$\eqalign{a=y+u\,,\;
\Delta={3\over 8x}\,,\;&\gamma={3\over 8v}\,,\;x+v=\h(y+u)^2\,,\cr
\lambda = (y+u)^{3/2}f(t)\,,&\quad it+1\equiv 4v/(y+u)^2\,,\cr 
\hbox{with~~~} 16(1+t^2)f'' + 3f = 0\;,&
\hbox{~$f$ a hypergeometric function.}\cr}\eqno(5)$$
In these coordinates its Killing vector is ${\widetilde K} = 
\p_u - \p_y$,
and its homothetic vector is given by ${\widetilde H} = 3p\,\p_p 
+ y\,\p_y + u\,\p_u + 2v\,\p_v$. 
\vs
Returning {\bf to our general form}, we also
choose a null tetrad, and specify the associated non-zero 
components of the curvature:
$$\eqalign{
{\bf g} = \dom^1\otimes\dom^2 + \dom^2\otimes
\dom^1 + \dom^3&\otimes\dom^4+\dom^4\otimes\dom^3\,,\quad\hbox{~with~}
\dom^1\ \equiv\ p\,du\,,\;\cr
\dom^2\ \equiv\ Z\,dy + a_1\,\dom^3\,,\;
\dom^3\ &\equiv dv - a\,du\,,\;
\dom^4\ \equiv\ dp + E\,du -  Q\,\dom^3\;,\cr
\hbox{where~}E \equiv \lambda(\lambda\,a_{32}+2\lambda_3\,a_2)\,,\;
Z &\equiv p/\lambda^2 + a_2\,,\;
Q \equiv p/\lambda^2 + \lambda^2(\lambda_2/\lambda)_3\;,\cr
\hbox{and~~}2 R_{1313}=2\gamma_1/p = C^{(1)}\;,&\quad
2 R_{2323} = 2(\lambda^2/Z)\Delta_1 = {\overline C}^{(1)}\;,
\cr}\eqno(6)$$
so that the curvature is indeed of type $N\otimes N$.
\vs
The forms of these constraining pde's are unchanged 
under any one of the following coordinate transformations.\cite{1}

\item{$\bf I.~$}Replace $\{v,y,u\}$ by $\{{\overline v},y,u\}$, 
with $~{\overline v}={\overline V}(v,u)$ 
arbitrary but invertible, along with 
$F$, $x$, $\gamma$, 
$\Delta$ scalars, while ${\overline\lambda} = \lambda/{\overline V}_v$
and ${\overline a} = {\overline V}_va + {\overline V}_u$;
\item{\bf II.~}Replace $\{x,y,u\}$ by $\{{\overline x},y,u\}$,
with $~{\overline x}={\overline X}(x,y)$
arbitrary but invertible, and $\lambda$, $a$, $v$, $\gamma$, 
$\Delta$ scalars, while ${\overline F} = F/{\overline X}_x$;
\item{\bf III.~}Replace $\{v,y,u\}$ by $\{v,{\overline y},u\}$,
with $~{\overline y}={\overline Y}({y})$ arbitrary but invertible, and $x$, $a$, $\gamma$ scalars,
while $\lambda$ [or $F$] scales as  
${\overline\lambda} = \sqrt{{\overline Y}_{y}}\;{\lambda}$, 
 and $\Delta$ has an additional term:~
$\Delta = {\overline\Delta} + \{\sqrt{{\overline Y}_y}\}_{yy}\;.$
\item{\bf IV.~}Replace $\{v,y,u\}$ by $\{v,y,{\overline u}\}$,
with $~u=U({\overline u})$ arbitrary but invertible, and
$x$, $\Delta$ scalars,
while $\lambda$ [or $F$] scales as  
${\overline\lambda} = \sqrt{{\overline U}_{u}}\;{\lambda}$, 
and also ${\overline a} = a/{\overline U}_u$, and $\gamma = {\overline\gamma} + \{\sqrt{{\overline U}_{u}}\}_{uu}\;.$
\par\no
We refer to $\gamma=\gamma(u,v)$ and $\Delta=\Delta(x,y)$ as gauge 
functions since transformations I and II would allow them
to be replaced by $v$ and $x$, respectively.  However, we 
save that freedom for now.

\ss{2}{Killing's Equations}

We reduce the generality of the pde's by insisting that 
the metric allow some symmetries. An arbitrary 
homothetic vector, $\widetilde V$, constrains the metric and curvature as follows:
$$
{\cal L}_{_{\tilde V}}\,g_{\alpha\beta} \equiv V_{(\alpha;\beta)} = 
2\chi_0\,g_{\alpha\beta}\,,\qquad
{\cal L}_{_{\tilde V}}\,{\dform{\Gamma}}^\alpha{}_{\beta}\ =\ 0\
= \ {\cal L}_{_{\tilde V}}\,{\dform{\Omega}}^\alpha{}_{\beta}\;.\eqno(7)$$
When put together with the pde's for the metric functions, 
Eqs.(3), via GRTensor and Maple, these constraints 
require any prospective homothetic vector to
be determined by only two functions, $K =  K(u)$ and  $B =  B(v,u)$:
$${\widetilde V} = +(2\chi_0 - B_{,v})p\p_p  
+{(\p_u+a\p_v - a_{,v})(B-aK)\over a_{,y}}\,\p_y
+ K\p_u + B\p_v\;,\eqno(8)$$
along with 
various constraints on $\lambda$, $a$, $\Delta$, and 
$\gamma$, relative to $K$ and $B$.
We may however use our coordinate freedom(s) to simplify those equations further:
$$\eqalign{\hbox{Under Transformation I,~~~}
{\overline v} = {\overline V}(v,u)\ \Longrightarrow\ &
{\overline K} = K\;,\quad {\overline B} = 
K{\overline V}_{,u} + B{\overline V}_{,v}\;;\cr
\hbox{under Transformation IV,~~~}{\overline u} = {\overline U}(u)
\ \Longrightarrow\ &{\overline K} = {\overline U}_{,u}\,K\;,\quad
{\overline B} = B\;.\cr}\eqno(9)$$
Therefore, when $K\ne 0$, we may always choose coordinates 
so that $B=0$ and $K$ is a constant, say $+1$, 
and then ask for the constraints on $\{\lambda,a,
\Delta,\gamma\}$ that are implied by this.
\vs  
\ss{3}{When one Homothetic Vector is Permitted}

With $K=+1$, $B=0$, Killing's equations require that 
$$\eqalign{
\gamma_{,v}=\gamma_{,v}(v)\,,\;\Delta_{,v}=&\ 
\Delta_{,v}[x(v,s)]\;,\;a=a(v,s)\,,\cr
\lambda=e^{\chi_{_{0}}\! u}L(v,s)\,,\;
\hbox{~ where~~} &s\equiv y+u \qquad
\hbox{and~~} 
 {\widetilde V} = 2\chi\nought p\,\p_p + \p_u - \p_y\;.\cr}
\eqno(10)$$
\vs\no
When $\chi\nought = 0$ this is the usual Killing vector,
namely a translation in the 2-plane which is the local wavefront. 
This clearly reduces the independent variables 
in the pde's to only two, so that the first 3 equations determine 
all the second derivatives of $\lambda$, and requiring 
only one more equation, 
solved for a 4th derivative of $a$, 
to complete the involutive set.
The new version of the constraining 
equations may be written in terms of only $\p_1$ and 
$\p_2$, or 
 in terms of only $\p_2$ and $\p_3$, with 
$\p_1\ \rightarrow\  \tfrac{1}{a}(\p_3-\p_2)$.\cite{6}
\vs
\ss{4}{One Homothetic Vector plus the Killing Vector}

We now indeed insist that our metric allows one true Killing 
vector, in the form described above for the metric quantities, with 
$\chi\nought =0$.\cite7  In addition we also 
ask for a second (homothetic) symmetry, 
${\widetilde H}$, which will have the form shown in Eqs.({8})
 with its own $K$, $B$, and 
$\chi\nought$ not necessarily zero.  Its existence is additionally
constrained by the fact that the commutator of two 
homothetic vectors must be a Killing vector:\cite{8} 
$$[{\widetilde K},{\widetilde H}] \propto 
{\widetilde K}\qquad\Longrightarrow\quad \p_uB = 0 = \p_u^2K
\eqno(11)$$
Since we have used some gauge freedom to simplify our Killing 
vector, much less freedom remains.  Nonetheless, while maintaining
the simple form of our (first) Killing vector, that freedom is still 
sufficient to allow us to set 
$$\eqalign{K\ \longrightarrow \ u\;,\quad\hbox{and}&\quad
B\ \longrightarrow\ 2v\,\cr \Longrightarrow\quad
{\widetilde H} = &\ 2(\chi\nought  - 1)p\,\p_p + y\,\p_y + u\,\p_u + 
2\, v\,\p_v\;,\cr}\eqno(12)$$
where the constant 2 is simply a convenient choice. 
\vs
 Killing's equations are then completely satisfied by 
the following ``scaling'' equations for each dependent variable
and the concommitant ones for their derivatives:
$$\eqalign{ {\widetilde H}(a) = a\;, &\qquad
 {\widetilde H}(\lambda) = 
(\chi\nought  - 1)\,\lambda\;,\cr
{\widetilde H}(\gamma) = -2\,\gamma\;,\qquad&\qquad 
 {\widetilde H}(\Delta) = -2\,\Delta\;.\cr}\eqno(13)$$
Since $a$ and $\lambda$ already depend only on $v$ and $s\equiv y+u$, 
these constraints reduce them in terms of functions of only 
one variable:
$$a = s\,A(q)\;,\qquad \lambda = 
s^{\chi\nought-1}L(q)\;,\qquad
q\ \equiv \ {v\over s^2}\;,\eqno(14)$$
\vskip-8pt\no
while the gauge functions are almost completely determined:
$$
\gamma = \gamma\nought/v = s^{-2}G(q), \hbox{~~i.e.,~}
 G = \gamma\nought/q\;,\quad
\Delta = s^{-2}D(q), \hbox{~where~~}
(A - 2q)(\ln D)^\prime = 2\;.\eqno(15)$$
\vs
Of course
 the original constraint equations must still be resolved.  They
are now 3 {\bf ode's} for the two functions, $L$ and $A$.
To display them, we take a new form for the 
similarity variable, 
$$ r \equiv \h\ln(q) \ \Leftrightarrow\ q
\equiv e^{2 r}\hbox{~~and set~~} W = W(r) \equiv (A-2 q)/(2 q) \;.
\eqno(16) $$
The first two equations present very nicely, in a 
simple, factorized form: 
$$\eqalign{
%\openup2\jot 
({d\over dr} + \chi\nought-1)
({d\over dr} + \chi\nought-2)L &\ = 
D\,L = e^{+2\!\int \!\!dr(1/W)}\,L\;,\cr\cr
({d\over dr}W + \chi\nought +1)({d\over dr} W+ \chi\nought)L & \ = 
GL = e^{-2p}L\;.\cr}\eqno(17)$$
\vs
\no
The third one, while still linear in $L$, is somewhat 
more complicated, and not immediately factorizable:
$$\eqalign{4WZ&\,{d^2L\over dr^2} + 2\left[({\bf A}+{\bf B})(WZ) + (2\chi\nought-1)Z\right]{dL\over dr} \cr 
 &\hskip1.0truein +\left\{{\bf A}({\bf B}(WZ)) + 
2(\chi\nought Z_r -\eta\nought\,Z)\right\}L = 0\;,\cr\cr
&\hbox{with~~} {\bf A} \equiv {d\over dr} + 4\,,
\quad {\bf B} \equiv 
{d\over dr} - (2\chi\nought - 5)\;,\cr
&\quad Z \equiv {dW\over dr} + W +1 , 
\hbox{~and~~} \eta\nought \equiv 2\chi\nought^2
-6\chi\nought +1
\;.\cr}\eqno(18)$$
\par
 Not all three equations, for two functions, are necessary.
Involutivity now needs only two of these equations. A possible 
choice is the following pair:
\itemitem{a.)~} a Riccati equation for $H\equiv L_{,r}/L$,
which is derived by using $\Delta_3=0$ to eliminate it 
from the presentation:
$$\eqalign{W[dW/dr&+(2\chi-1)(W+1)]{dH\over dr}\cr
& - 2W[dW/dr+W+1]H^2 + \mu H + \nu = 0\;,\cr}\eqno(19)$$
where $\mu$ and $\nu$ are fairly complicated polynomials in 
 $(WW_r)_r$, $W_r$, $W$, and constants; and 
\itemitem{b.)~} the equation above, Eq. (17b), 
with $W$ and $G$, which is second order and linear for $L$,
but which could also be seen as a Riccati equation for $H$.

\veject
%%This is a page for references!
\centerline{REFERENCES}

\parindent=0pt
\baselineskip=12pt
\def\hi{\hangindent=20truept}

\frenchspacing
\font\itabs=cmti9
\def\JMP{{\itabs J. Math. Phys.~}}
\def\CQG{{\itabs Cl. Qu. Grav.~}}

\vs
\smaller

\hi$^\dagger$ From a talk presented at the 16th International Conference
of the Society on General Relativity and Gravitation, 15-21 
July, 2001, Durban, South Africa
\vs
\hi 1. J.D. Finley, III and J.F. Pleba\'nski, 
``Equations for twisting, 
type-N, vacuum Einstein spaces without a need for Killing vectors,"
J. Geom. Phys. {\smbold 8} 173-193 (1992).

\hi 2.  
Other research in this area includes C.B.G. McIntosh, ``Symmetries of
vacuum type-N metrics," \CQG{\smbold 2}87-97 (1985), and 
``Twisting Type N Vacuum Solutions which admit an $H_2$ of 
Homothetic Killing Fields," presented at the Fifth Hungarian 
Relativity Workshop, Budapest, August, 1995; 
H. Stephani and E. Herlt, ``Twisting type-N vacuum solutions with two
non-commuting Killing vectors do exist," 
\CQG {\smbold 2} L63-64 (1985); F.J. Chinea, ``New first integral for 
twisting type-N vacuum gravitational fields with two non-commuting 
Killing vectors,"
\CQG {\smbold 15} 367-371 (1998); and J.D. Finley, III, 
J.F. Pleba\'nski and Maciej Przanowski, ``Third-order ode's for 
twisting type-N vacuum solutions," 
\CQG {\smbold 11} 157-166 (1994).

\hi 3.
 C.P. Boyer, J.D. Finley, III and J.F. Pleba\'nski, ``Complex 
general relativity and \hsp\hsp spaces---A survey of one approach," in 
{\itabs General Relativity and Gravitation,} Vol. 2, A. Held (Ed.) (Plenum,
New York, 1980) pp. 241-281.

\hi 4. J.D. Finley, III and Andrew Price, 
``The Involutive Prolongation of 
the (Complex) Twisting, Type-N Vacuum Field Equations," in 
{\itabs Aspects of 
General Relativity and mathematical Physics (Proceedings of a Conference in 
Honor of Jerzy Pleba\'nski)}, Nora Bret\'on, Riccardo Capovilla \&
Tonatiuh Matos (Eds.) (Centro de Investigaci\'on y de Estudios Avanzados 
del I.P.N., Mexico City, 1993).

\hi 5.  
I. Hauser, ``Type N gravitational field with twist," {\itabs Phys. 
Rev. Lett.}~ {\smbold 33} 1112-1113 (1974), and \JMP {\smbold 19}
 661-667 (1978).

\hi 6.  D. Khetselius, ``Nonlocal Prolongations for the 
Twisting Type-N, Einstein Vacuum Field Equations with One 
Killing Vector," (Ph.D. Dissertation, Univ. New Mexico, 1996,
unpublished)

\hi 7.  This form for the equations was accomplished working with 
 Francisco Navarro Lerida, U. Compl. de Madrid.

\hi 8.  Theorem cited in W.D. Halford, ``Einstein spaces and 
homothetic motions. II,'' \JMP {\smbold 21} 129-134 (1980); 
original reference in K. Yano, J. Indian Math. Soc. {\smbold 
15} 105 (1951).

\bye